\documentclass[11pt, a4paper]{article}
\usepackage{graphicx}
\usepackage{amssymb}
\usepackage{amsmath}
\usepackage{bm}
\usepackage{color}
\usepackage{theorem}
\usepackage{subcaption}

\usepackage[sort&compress,numbers, merge]{natbib}

\definecolor{Orange}{cmyk}{0,0.61,0.87,0}
\definecolor{JungleGreen}{cmyk}{0.99,0,0.52,0}
\definecolor{OliveGreen}{cmyk}{0.64,0,0.95,0.40}
\definecolor{Brown}{cmyk}{0,0.81,1,0.60}
\definecolor{RoyalBlue}{cmyk}{0.71,0.53,0,0.12}

\usepackage[colorlinks=true, linkcolor=OliveGreen, citecolor=RoyalBlue, urlcolor=Brown]{hyperref}

\begin{document}

\begin{titlepage}

\begin{flushright}
{\tt 
UT-17-17\\
IPMU-17-0072
}
\end{flushright}

\vskip 1.35cm
\begin{center}

{\Large
{\bf
Predictions for the neutrino parameters\\[3pt]
in the minimal gauged U(1)$_{L_\mu-L_\tau}$ model
}
}

\vskip 1.2cm

Kento Asai$^{1}$, 
Koichi Hamaguchi$^{1,2}$, 
Natsumi Nagata$^{1}$

\vskip 0.4cm

{\it $^1$Department of Physics, University of Tokyo, Bunkyo-ku, Tokyo
 133--0033, Japan} \\[2pt]
{\it ${}^2$Kavli Institute for the Physics and Mathematics of the
 Universe (Kavli IPMU), University of Tokyo, Kashiwa 277--8583, Japan} 

\date{\today}

\vskip 1.5cm

\begin{abstract}
 We study the structure of the neutrino mass matrix in the minimal gauged
 U(1)$_{L_\mu-L_\tau}$ model, where three right-handed neutrinos are
 added to the Standard Model in order to obtain non-zero masses for
 active neutrinos. Because of the U(1)$_{L_\mu-L_\tau}$ gauge symmetry,
 the structure of both Dirac and Majorana mass terms of neutrinos is
 tightly restricted. In particular, the {\it inverse} of the neutrino
 mass matrix has zeros in the $(\mu,\mu)$ and $(\tau,\tau)$ components,
 namely, this model offers a symmetric realization of the so-called
 two-zero-minor structure in the neutrino mass matrix. Due to these
 constraints, all the CP phases---the Dirac CP phase $\delta$ and
 the Majorana CP phases $\alpha_2$ and $\alpha_3$---as well as the mass
 eigenvalues of the light neutrinos $m_i$ are uniquely determined as
 functions of the neutrino mixing angles $\theta_{12}$, $\theta_{23}$,
 and $\theta_{13}$, and the squared mass differences $\Delta m_{21}^2$
 and $\Delta m_{31}^2$. We find that this model predicts the Dirac CP
 phase $\delta$ to be $\delta\simeq 1.59\pi$--$1.70\pi$
 ($1.54\pi$--$1.78\pi$), the sum of the neutrino masses to be
 $\sum_{i}m_i \simeq 0.14$--0.22~eV ($0.12$--0.40~eV), and the
 effective mass for the neutrinoless double beta decay to be $\langle
 m_{\beta \beta }\rangle \simeq 0.024$--0.055~eV ($0.017$--0.12~eV) at
 $1\sigma$ ($2\sigma$) level, which are totally consistent with the
 current experimental limits. These predictions can soon be tested in
 future neutrino experiments. 
 Implications for leptogenesis are also discussed.

\end{abstract}

\end{center}
\end{titlepage}

\section{Introduction}
\label{sec:intro}

A gauged U(1)$_{L_\mu-L_\tau}$ symmetric extension of the Standard Model
(SM), where $L_\mu$ and $L_\tau$ stand for the $\mu$ and $\tau$ numbers,
respectively, has widely been considered so far, as it is one of the several
possibilities of gauging an accidental U(1) symmetry in the SM
\cite{Foot:1990mn, He:1990pn, He:1991qd, Foot:1994vd}. An
attractive feature of this class of models is that the muon $g-2$
anomaly \cite{Bennett:2006fi, Jegerlehner:2009ry, Davier:2010nc,
Hagiwara:2011af} may be explained by the loop contribution of the
U(1)$_{L_\mu-L_\tau}$ gauge boson if its mass lies around the weak scale
or lower \cite{Baek:2001kca, Ma:2001md, Heeck:2011wj, Harigaya:2013twa},
though this possibility is severely restricted by the searches of the
neutrino trident production process \cite{Geiregat:1990gz,
Mishra:1991bv, Altmannshofer:2014cfa, Altmannshofer:2014pba}. 
These models may also explain anomalies in flavor physics
\cite{Altmannshofer:2014cfa, Crivellin:2015mga}
and offer promising candidates for dark matter in the Universe
\cite{Kim:2015fpa, Baek:2015fea, Patra:2016shz, Biswas:2016yan,
Biswas:2016yjr}. For other recent studies on gauged
U(1)$_{L_\mu-L_\tau}$ models, see
Refs.~\cite{delAguila:2014soa, Fuyuto:2014cya, Baek:2015mna, Araki:2015mya,
Elahi:2015vzh, Fuyuto:2015gmk, Altmannshofer:2016oaq, Ibe:2016dir,
Kaneta:2016uyt, Araki:2017wyg, Lee:2017ekw, Hou:2017ozb, Chen:2017cic}. 

On the other hand, neutrino oscillation data show that at least
two of the active neutrinos have non-zero masses and there is sizable
mixing among these neutrinos. This can be accounted for if we add 
right-handed neutrinos to the theory and couple them to the SM lepton
and Higgs fields through the Yukawa couplings. After the
Higgs field acquires a vacuum expectation value (VEV), these terms lead
to the Dirac mass terms for the neutrinos. In addition, the right-handed
neutrinos can have Majorana mass terms, and if these masses are much
larger than the electroweak scale, then the smallness of the neutrino
masses can naturally be explained by the seesaw mechanism
\cite{Minkowski:1977sc, Yanagida:1979as, GellMann:1980vs, Mohapatra:1979ia}. 

In gauged U(1)$_{L_\mu-L_\tau}$ models, however, the neutrino mass
structure is tightly restricted since the second and third generation
leptons are charged under the U(1)$_{L_\mu-L_\tau}$ gauge symmetry
\cite{Branco:1988ex, Choubey:2004hn, Araki:2012ip, Heeck:2014sna,
Plestid:2016esp}. In fact, in the U(1)$_{L_\mu-L_\tau}$ preserving
limit, the Dirac mass matrix is diagonal, while in the Majorana mass
matrix only the $(e, e)$, $(\mu, \tau)$, and $(\tau, \mu)$ components
can be non-zero. Such a simple structure cannot explain neutrino
oscillation data, and therefore we need to break the
U(1)$_{L_\mu-L_\tau}$ symmetry. To that end, one usually introduces a SM
singlet scalar that has a non-zero U(1)$_{L_\mu-L_\tau}$ charge such
that its VEV spontaneously breaks the U(1)$_{L_\mu-L_\tau}$ gauge
symmetry and gives a mass to the U(1)$_{L_\mu-L_\tau}$ gauge boson. We
however note that even though the U(1)$_{L_\mu-L_\tau}$ gauge symmetry
is spontaneously broken, the structure of the neutrino mass matrix is
still highly constrained if we introduce just one
U(1)$_{L_\mu-L_\tau}$-breaking scalar field and consider only
renormalizable interactions. Therefore, it is interesting to study if
this minimal setup can accommodate the neutrino mass structure that is
consistent with the present neutrino oscillation data.

In this paper, we consider this ``minimal'' gauged U(1)$_{L_\mu-L_\tau}$
model where three right-handed neutrinos and one 
U(1)$_{L_\mu-L_\tau}$-charged SM singlet scalar field are added to the SM.
Then, it turns out that the observed neutrino mixing structure can be
obtained only when the U(1)$_{L_\mu-L_\tau}$-breaking scalar field has
the U(1)$_{L_\mu-L_\tau}$ charge $\pm 1$. In this case, the $(\mu,\mu)$ and
$(\tau,\tau)$ components of the Majorana mass matrix for the
right-handed neutrinos remain zero even after the U(1)$_{L_\mu-L_\tau}$
symmetry is spontaneously broken. Because of this structure of the
Majorana mass matrix together with the diagonal Dirac mass matrix, the
{\it inverse} of the neutrino mass matrix also has zeros in the
$(\mu,\mu)$ and $(\tau,\tau)$ components \cite{Araki:2012ip,Crivellin:2015lwa}. The minimal
gauged U(1)$_{L_\mu-L_\tau}$ model thus gives a concrete realization of
a two-zero-minor model \cite{Lavoura:2004tu,
Lashin:2007dm}. Intriguingly, due to the condition that the $(\mu,\mu)$
and $(\tau,\tau)$ components in the inverse of the neutrino mass matrix
vanish, all the CP phases in the neutrino mixing matrix---the Dirac CP
phase $\delta$ and the Majorana CP phases $\alpha_2$ and $\alpha_3$---as
well as the mass eigenvalues of the light neutrinos are
uniquely determined as functions of the neutrino mixing angles
$\theta_{12}$, $\theta_{23}$, and $\theta_{13}$, and the squared mass
differences $\Delta m_{21}^2$ and $\Delta m_{31}^2$. As we see below,
this prediction is independent of the U(1)$_{L_\mu-L_\tau}$-symmetry
breaking scale, and thus can be regarded as a generic prediction in the
minimal gauged U(1)$_{L_\mu-L_\tau}$ model. We find that the predicted
values of the neutrino parameters are consistent with the present
neutrino data and can be tested in future neutrino experiments. 
We also discuss the implications of our results for leptogenesis.

This paper is organized as follows. In Sec.~\ref{sec:model}, we
introduce the minimal gauged U(1)$_{L_\mu-L_\tau}$ model and examine the
neutrino mass structure in this model. We then show in
Sec.~\ref{sec:result} the predicted values of the Dirac CP phase
$\delta$, the sum of the neutrino masses $\sum_{i}m_i$, and the
effective mass for the neutrinoless double beta decay $\langle m_{\beta
\beta} \rangle$ using the neutrino mixing angles and the squared mass
differences obtained in neutrino oscillation experiments. 
In Sec.~\ref{sec:LG}, we discuss the implications for the leptogenesis.
Finally, our
conclusions are summarized in Sec.~\ref{sec:conclusion}. This paper ends
with two Appendices which  
give further details of our analysis. In Appendix~\ref{sec:misc}, we
present some miscellaneous formulae which are useful to study the
neutrino structure in the minimal gauged U(1)$_{L_\mu-L_\tau}$ model. In
Appendix~\ref{sec:othercases}, we perform similar analyses for the
minimal gauged  U(1)$_{L_e-L_\mu}$ and U(1)$_{L_e-L_\tau}$ models, and
show that these minimal models fail to explain the observed neutrino
oscillation data.

\section{Neutrino mass structure in the minimal gauged
 U(1)$_{L_\mu-L_\tau}$ model} 
\label{sec:model}

To begin with, we describe the minimal U(1)$_{L_\mu-L_\tau}$ model which
we discuss in this paper. The model possesses a new U(1) gauge
symmetry U(1)$_{L_\mu - L_\tau}$. Under this gauge symmetry, $\mu_{L,R}$
and $\nu_\mu$ have the U(1)$_{L_\mu - L_\tau}$ charge $+1$, $\tau_{L,
R}$ and $\nu_\tau$ have the U(1)$_{L_\mu - L_\tau}$ charge $-1$, and the
other SM fields have the zero U(1)$_{L_\mu - L_\tau}$ charge. We also
introduce three right-handed neutrinos $N_e$, $N_\mu$, and $N_\tau$ to
obtain non-zero neutrino masses. The U(1)$_{L_\mu - L_\tau}$ charges of
these fields are $0$, $+1$, and $-1$, respectively.\footnote{Generically
speaking, the U(1)$_{L_\mu - L_\tau}$ charges of the right-handed
neutrinos can be $(0, a, -a)$ without spoiling anomaly cancellation
conditions. We however find that only the $|a|=1$ case gives a neutrino
mass structure that can explain the neutrino oscillations. Then we can
define the right-handed neutrinos with the U(1)$_{L_\mu - L_\tau}$
charge 0, $+1$, and $-1$, as $N_e$, $N_\mu$, and $N_\tau$, respectively,
without loss of generality. We also note that the introduction of two
right-handed neutrinos is insufficient since the observed neutrino
mixing angles cannot be reproduced in this case.  
} 

In the $(e, \mu, \tau)$ basis, the U(1)$_{L_\mu - L_\tau}$
charges of the Dirac Yukawa terms are
\begin{equation}
Q_{L_\mu - L_\tau}(\text{Dirac}):
~~
 \begin{pmatrix}
  0&1&-1 \\
  -1& 0 & -2 \\
  1 & 2 & 0
 \end{pmatrix}
~,
\label{eq:charged}
\end{equation}
where the $(\alpha, \beta)$ entry in the above matrix represents the
U(1)$_{L_\mu - L_\tau}$ charge of the fermion bilinear term $N_\alpha^c
L_\beta^{}$, with $\alpha, \beta$ the flavor indices. For the
Majorana mass terms of the right-handed neutrinos, on the other hand, we
have  
\begin{equation}
Q_{L_\mu - L_\tau}(\text{Majorana}):
~~
 \begin{pmatrix}
  0 & 1 & -1 \\
  1 & 2 & 0 \\
  -1& 0 & -2 \\
 \end{pmatrix}
~,
\label{eq:chargem}
\end{equation}
where the $(\alpha, \beta)$ component indicates the U(1)$_{L_\mu -
L_\tau}$ charge of the fermion bilinear term $N_\alpha N_\beta$.
From Eq.~\eqref{eq:charged}, we find that the Dirac Yukawa matrix is
always diagonal in the gauged U(1)$_{L_\mu - L_\tau}$ models---for the
same reason, the charged lepton Yukawa matrix is also diagonal. As long as
renormalizable interactions are considered, this structure is not
violated even if we introduce a U(1)$_{L_\mu - L_\tau}$-breaking scalar
field. On the other hand, the charges of the Majorana mass matrix in
Eq.~\eqref{eq:chargem} show that only the $(e,e)$, $(\mu,\tau)$, and
$(\tau, \mu)$ components can have non-zero values in the U(1)$_{L_\mu -
L_\tau}$-symmetric limit. As we mentioned above, with this simple
structure, we cannot explain the required values of the neutrino mixing
angles. We therefore introduce a scalar boson
$\sigma$ which has a non-zero U(1)$_{L_\mu - L_\tau}$ charge, and couple
it to right-handed neutrinos. 
After this scalar field develops a VEV, these couplings lead to
Majorana mass terms of the right-handed neutrinos. If the scalar field
has the U(1)$_{L_\mu - L_\tau}$ charge $\pm 1$, then the $(e,\mu)$, $(e, 
\tau)$, $(\mu,e)$, and $(\tau, e)$ components in Eq.~\eqref{eq:chargem}
can be induced after the scalar field acquires a VEV, while the $(\mu,
\mu)$ and $(\tau, \tau)$ can be generated if the scalar has the
U(1)$_{L_\mu - L_\tau}$ charge $\pm 2$. In the latter case, however, the
Majorana mass matrix becomes block-diagonal, which makes it unable to explain
the observed neutrino mixing angles. We are thus left with the case
where the scalar field has the U(1)$_{L_\mu - L_\tau}$ charge $\pm 1$,
and we take it to be $+1$ in the following discussion. We refer to this
model as the minimal gauged U(1)$_{L_\mu-L_\tau}$ model.

The interaction terms relevant to neutrino masses are then given by
\begin{align}
 \Delta {\cal L} = 
&-\lambda_e N_e^c (L_e \cdot H)
-\lambda_\mu N_\mu^c (L_\mu \cdot H)
-\lambda_\tau N_\tau^c (L_\tau \cdot H) \nonumber \\
&-\frac{1}{2}M_{ee} N_e^c N_e^c 
- M_{\mu \tau} N_\mu^c N_\tau^c 
- \lambda_{e\mu} \sigma N_e^c N_\mu^c
- \lambda_{e\tau} \sigma^* N_e^c N_\tau^c +\text{h.c.} ~,
\label{eq:lag}
\end{align}
where the dots indicate the contraction of the SU(2)$_L$ indices.
After the Higgs field $H$ and the singlet scalar $\sigma$ acquire VEVs
$\langle H \rangle = v/\sqrt{2}$ and $\langle \sigma
\rangle$,\footnote{We can always take the VEV of $\sigma$ to be real by
using U(1)$_{L_\mu - L_\tau}$ transformations. }
respectively, these interaction terms lead to the neutrino mass terms:
\begin{align}
 {\cal L}_{\rm mass} &= -(\nu_e, \nu_\mu, \nu_\tau) {\cal M}_D 
\begin{pmatrix}
N_e^c\\ N_\mu^c\\ N_\tau^c 
\end{pmatrix}
- \frac{1}{2}(N_e^c, N_\mu^c, N_\tau^c) {\cal M}_R 
\begin{pmatrix}
 N_e^c \\ N_\mu^c \\ N_\tau^c 
\end{pmatrix}
+\text{h.c.} ~,
\end{align}
where
\begin{equation}
 {\cal M}_D = \frac{v}{\sqrt{2}}
\begin{pmatrix}
 \lambda_e & 0& 0\\
 0 & \lambda_\mu & 0 \\
 0 & 0 & \lambda_\tau 
\end{pmatrix}
~,\qquad
{\cal M}_R =
\begin{pmatrix}
 M_{ee} & \lambda_{e\mu} \langle \sigma \rangle & \lambda_{e\tau} 
\langle \sigma \rangle \\
 \lambda_{e\mu} \langle \sigma \rangle & 0 & M_{\mu\tau} \\
\lambda_{e\tau} \langle \sigma \rangle & M_{\mu\tau} & 0
\end{pmatrix}
~.
\end{equation}
The mass matrix for the light neutrinos is then given by
\cite{Minkowski:1977sc, Yanagida:1979as, GellMann:1980vs,
Mohapatra:1979ia} 
\begin{equation}
 {\cal M}_{\nu_L} \simeq - {\cal M}_D {\cal M}_R^{-1} {\cal M}_D^T ~.
\label{eq:mnul}
\end{equation}
An explicit expression for ${\cal M}_{\nu_L}$ can be found in
Appendix~\ref{app:mnul}. 
We can diagonalize this mass matrix by using a unitary matrix $U$ (PMNS
matrix~\cite{Olive:2016xmw}):  
\begin{equation}
 U^T {\cal M}_{\nu_L} U =\text{diag}(m_1, m_2, m_3) ~,
\label{eq:diagmnul}
\end{equation}
which can be parametrized as 
\begin{equation}
 U = 
\begin{pmatrix}
 c_{12} c_{13} & s_{12} c_{13} & s_{13} e^{-i\delta} \\
 -s_{12} c_{23} -c_{12} s_{23} s_{13} e^{i\delta}
& c_{12} c_{23} -s_{12} s_{23} s_{13} e^{i\delta}
& s_{23} c_{13}\\
s_{12} s_{23} -c_{12} c_{23} s_{13} e^{i\delta}
& -c_{12} s_{23} -s_{12} c_{23} s_{13} e^{i\delta}
& c_{23} c_{13}
\end{pmatrix}
\begin{pmatrix}
 1 & & \\
 & e^{i\frac{\alpha_{2}}{2}} & \\
 & & e^{i\frac{\alpha_{3}}{2}}
\end{pmatrix}
~,
\end{equation}
where $c_{ij} \equiv \cos \theta_{ij}$ and $s_{ij} \equiv \sin
\theta_{ij}$ for $\theta_{ij} = [0, \pi/2]$, $\delta = [0, 2\pi]$,
and we have ordered $m_1<m_2$ without loss of generality. 
We follow the convention of the Particle Data Group~\cite{Olive:2016xmw}, 
where $m_2^2-m_1^2\ll |m_3^2-m_1^2|$ and $m_1<m_2<m_3$ (Normal Ordering,
NO) or $m_3<m_1<m_2$ (Inverted Ordering, IO).

If $m_i = 0$ ($i = 1$ or 3), then $\text{det}({\cal M}_{\nu_L}) =
0$. As shown in Appendix~\ref{app:mnul}, in this case we cannot have
desired mixing angles since ${\cal M}_{\nu_L}$ becomes
block-diagonal. Thus, we focus on the $m_i \neq 0$ case, where we obtain
from Eqs.~\eqref{eq:mnul} and \eqref{eq:diagmnul} 
\begin{align}
 {\cal M}_{\nu_L}^{-1} = U  \text{diag}(m_1^{-1}, m_2^{-1}, m_3^{-1})
 U^T
\simeq - ({\cal M}_D^{-1})^T {\cal M}_R {\cal M}_D^{-1} ~.
\end{align}
We then notice that the $(\mu, \mu)$ and $(\tau, \tau)$ components of
these terms vanish since ${\cal M}_D$ is diagonal and ${\cal M}_R$ has
zeros in these components.  This structure is sometimes called two-zero
minor \cite{Lavoura:2004tu, Lashin:2007dm}. For other previous studies
of the two-zero minor structure, see Refs.~\cite{Verma:2011kz,
Liao:2013saa}. These two vanishing conditions then lead to 
\begin{align}
 \frac{1}{m_1}V_{\mu 1}^2 + \frac{1}{m_2}V_{\mu 2}^2\,e^{i\alpha_2}
+ \frac{1}{m_3}V_{\mu 3}^2\,e^{i\alpha_3} &= 0 ~,
\label{eq:vanmu}\\[3pt]
 \frac{1}{m_1}V_{\tau 1}^2 + \frac{1}{m_2}V_{\tau 2}^2\,e^{i\alpha_2}
+ \frac{1}{m_3}V_{\tau 3}^2\,e^{i\alpha_3} &= 0 ~,
\label{eq:vantau}
\end{align}
where the matrix $V$ is defined by $U = V\cdot \text{diag}(1,
e^{i\alpha_2/2}, e^{i\alpha_3/2})$. Notice that neither the U(1)$_{L_\mu
- L_\tau}$-breaking singlet VEV $\langle \sigma \rangle$ nor Majorana
masses $M_{ee}$ and $M_{\mu\tau}$ appear in these conditions
explicitly. For this reason, the following discussions based on these
equations have little dependence on the U(1)$_{L_\mu - L_\tau}$-symmetry
breaking scale; it may be around the electroweak scale, or as large as
$10^{(13-15)}$~GeV, which is a prime scale for the masses of
right-handed neutrinos since small neutrino masses are explained with 
${\cal O}(1)$ Yukawa couplings via the seesaw mechanism
\cite{Minkowski:1977sc, Yanagida:1979as, GellMann:1980vs,
Mohapatra:1979ia}. It follows from Eqs.~\eqref{eq:vanmu} and
\eqref{eq:vantau} that 
\begin{align}
 e^{i\alpha_2} &=\frac{m_2}{m_1} R_2 (\delta) ~,
\qquad
 e^{i\alpha_3} =\frac{m_3}{m_1} R_3 (\delta) ~,
\label{eq:alp23}
\end{align}
with\footnote{These expressions are consistent with the results
presented in Ref.~\cite{Liao:2013saa}. We also find from the explicit
expressions \eqref{eq:r2ex} and \eqref{eq:r3ex} that the corresponding
equations in Ref.~\cite{Verma:2011kz} disagree with ours.}
\begin{align}
 R_2 &\equiv \frac{(V_{\mu 1} V_{\tau 3} + V_{\mu 3} V_{\tau 1})V^*_{e2}}
{(V_{\mu 2} V_{\tau 3}+ V_{\mu 3} V_{\tau 2}) V^*_{e1}} ~, 
\label{eq:r2}\\
 R_3 &\equiv \frac{(V_{\mu 1} V_{\tau 2} + V_{\mu 2} V_{\tau 1})V^*_{e3}}
{(V_{\mu 2} V_{\tau 3}+ V_{\mu 3} V_{\tau 2}) V^*_{e1}} ~,
\label{eq:r3}
\end{align}
where we have used $\widetilde{V}^T = V^{-1}$ and ${\rm det}V=1$ with
$\widetilde{V}$ being the cofactor matrix of $V$.\footnote{The cofactor
$\tilde{A}_{ij}$ of a matrix $A$ is given by the determinant of the submatrix
formed by removing the $i$-th row and $j$-th column of the matrix $A$,
multiplied by a factor of $(-1)^{i+j}$. The cofactor matrix of $A$,
$\widetilde{A}$, is defined by $\widetilde{A} \equiv (\tilde{A}_{ij})$. We then
have ${A}^{-1} = ({\rm det} A)^{-1} \widetilde{A}^T$. }
In Appendix~\ref{app:r2r3}, we give explicit expressions for $R_2$ and
$R_3$ in terms of neutrino oscillation parameters. By taking the
absolute values of the equations in \eqref{eq:alp23}, we find
\begin{equation}
 \frac{m_2}{m_1} = \frac{1}{|R_2(\delta)|} ~,\qquad
 \frac{m_3}{m_1} = \frac{1}{|R_3(\delta)|} ~.
\label{eq:m2m3}
\end{equation}
Therefore, these mass ratios are given as functions of the Dirac CP phase
$\delta$. Notice that $R_{2,3}^*(-\delta) = R_{2,3}$ since $R_{2,3}$
contains a single CP phase $\delta$. As a consequence, $m_{2,3}/m_1$ are
symmetric under the reflection $\delta \to -\delta$ (and $\pi + \delta
\to \pi - \delta$) as we see below. 

\begin{figure}[t]
\centering
\includegraphics[clip, width = 0.5 \textwidth]{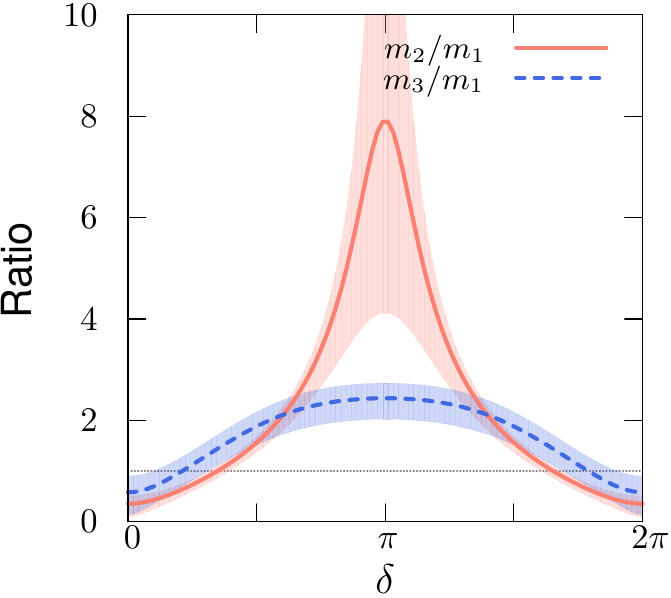}
\caption{The mass ratios $m_2/m_1$ and $m_3/m_1$ as functions of the
 Dirac CP phase $\delta$, from Eq.~\eqref{eq:m2m3}. The bands show
 uncertainty coming from the 1$\sigma$ error in the neutrino mixing
 parameters. The thin dotted line corresponds to $m_{2,3}/m_1 = 1$. }
\label{fig:ratio}
\end{figure}

In Fig.~\ref{fig:ratio}, we plot the mass ratios $m_2/m_1$ and $m_3/m_1$
against the Dirac CP phase $\delta$ using Eq.~\eqref{eq:m2m3}.  
The bands show uncertainty
coming from the 1$\sigma$ error in the neutrino mixing parameters. For
input parameters of the neutrino mixing angles, we use
$\sin^2\theta_{12} = 2.97^{+0.17}_{-0.16}\times 10^{-1}$, 
$\sin^2\theta_{13} = 2.15^{+0.07}_{-0.07}\times 10^{-2}$, and
$\sin^2\theta_{23} = 4.25^{+0.21}_{-0.15}\times 10^{-1}$
\cite{Capozzi:2017ipn} (see Table~\ref{tab:input}).\footnote{See also Refs.\cite{Forero:2014bxa,Esteban:2016qun}.} It is found that the
resultant uncertainty mainly comes from the error in $\theta_{23}$.
From this figure, we find that when $\delta \simeq 0,\,
2\pi$ the observed neutrino mixing angles are incompatible with the
condition  $m_1 < m_2$, and thus these regions are excluded. 
Moreover, around $\delta \simeq \pi$ we have $m_1< m_3 < m_2$, which
disagrees with the possible neutrino mass ordering: either $m_1<m_2<
m_3$ or $m_3<m_1<m_2$ \cite{Olive:2016xmw}. As a consequence, in the
region where a consistent neutrino mass ordering is obtained, the
neutrino mass ordering is always Quasi-Degenerate Normal Ordering with
$m_1\lesssim m_2 \lesssim m_3$, which is realized around $\delta\sim
\pi/2$ and $3\pi/2$.

According to the recent global-fit analysis in Ref.~\cite{Capozzi:2017ipn},
the NO is somewhat favored over IO at $\sim 2\sigma$ level. There are
quite a few proposed experiments that may determine the neutrino mass
ordering at more than 3$\sigma$ level within a decade \cite{Qian:2015waa,
Patterson:2015xja}, such as PINGU \cite{Aartsen:2014oha}, ORCA
\cite{Adrian-Martinez:2016fdl}, and JUNO \cite{An:2015jdp,
Djurcic:2015vqa}. To confirm NO in these future experiments would be
a first consistency check of our model.

Now that the mass ordering has been fixed, we determine $m_1$ and
$\delta$ by using the above results. From the neutrino oscillation
experiments, we can measure\footnote{
In terms of the squared mass
differences $\Delta m_{21}^2 \equiv m_2^2 -m_1^2$ and $\Delta m_{31}^2
\equiv m_3^2 -m_1^2$, $\delta m^2$ and $\Delta m^2$ are expressed as
\begin{equation}
 \delta m^2 = \Delta m_{21}^2~,\qquad
 \Delta m^2 = \Delta m_{31}^2 -\frac{1}{2}\Delta m_{21}^2 ~.
\end{equation}
} 
\begin{align}
 \delta m^2 &\equiv m_2^2 - m_1^2 ~, \\
 \Delta m^2 &\equiv m_3^2 -(m_2^2 + m_1^2)/2 ~.
\end{align}
These quantities are related to the neutrino mass ratios as
\begin{align}
 \delta m^2 &= m_1^2 \left(\frac{m_2^2}{m_1^2}-1 \right) 
= m_1^2\left(\frac{1}{\left|R_2(\delta)\right|^2}-1\right)
~,
\label{eq:delm2} 
\\[3pt]
 \Delta m^2 + \frac{\delta m^2}{2}
 &= m_1^2 \left(\frac{m_3^2}{m_1^2}-1 \right) 
= m_1^2\left(\frac{1}{\left|R_3(\delta)\right|^2}-1\right)
~.
\label{eq:Delm2}
\end{align}
By solving these equations, we can determine $m_1$ and $\delta$. 
The observed values of $\delta m^2$ and $\Delta m^2$ are
given by $\delta m^2 \simeq 7.37 \times 10^{-5}~\text{eV}^2$
and $\Delta m^2 \simeq 2.525 \times 10^{-3}~\text{eV}^2$,
respectively \cite{Capozzi:2017ipn}. This means that the right-hand side
of Eq.~\eqref{eq:delm2} is much smaller than that in
Eq.~\eqref{eq:Delm2}. From Fig.~\ref{fig:ratio}, we see that such a
hierarchy can be realized only when $m_2^2/m_1^2 \simeq 1$,
\textit{i.e.}, $|R_2 (\delta)| \simeq 1$. With the explicit formula of
$R_2(\delta)$ in Eq.~\eqref{eq:r2ex}, this leads to
\begin{equation}
 \cos \delta \simeq \frac{\cot 2\theta_{12} \cot 2\theta_{23}}{\sin
  \theta_{13}} ~.
\label{eq:appcosd}
\end{equation}
When the best-fit values of the mixing angles $\theta_{ij}$ are used,
this leads to $\cos \delta\simeq 0.46$, which corresponds to $\delta
\simeq 0.35\pi$ or $1.65\pi$. In Eq.~\eqref{eq:cubiceq} in
Appendix~\ref{sec:cub}, we give a cubic equation whose solution gives an
exact value of $\cos\delta$ as a function of the mixing angles
$\theta_{ij}$ and the squared mass differences $\delta m^2$ and $\Delta
m^2$. As discussed there, the solution \eqref{eq:appcosd}
approximates the real solution of the cubic equation \eqref{eq:cubiceq}
at ${\cal O}(\delta m^2/\Delta m^2)$ level. By solving
Eq.~\eqref{eq:cubiceq} numerically, we find $\cos \delta = 0.445$ for
the best-fit values of $\theta_{ij}$, $\delta m^2$, and $\Delta m^2$,
which means $\delta = 0.353\pi$ or 
$1.647\pi$, and justifies the expected accuracy of the approximated
formula \eqref{eq:appcosd}.\footnote{This result disagrees with the
observation in Ref.~\cite{Biswas:2016yan}, where 
it was observed that $\delta \simeq 0$ is predicted in the minimal
gauged $L_\mu-L_\tau$ model. In the analysis of
Ref.~\cite{Biswas:2016yan}, the Dirac Yukawa terms of
neutrinos are supposed to be real in the basis where ${\cal M}_R$ has
only one CP phase, though it is not the generic case. In addition, we
find disagreement between the neutrino mass matrix shown in
Ref.~\cite{Biswas:2016yan} and ours, as pointed out in
Appendix~\ref{app:mnul}.}

From Eqs.~\eqref{eq:r2} and \eqref{eq:delm2}, we obtain $m_1$
as a function of $\delta$. An explicit expression for $m_1$ is given in
Appendix~\ref{app:r2r3}. As noted above, $|R_{2,3} (\delta)|$ are
symmetric with respect to the reflection $\delta \to -\delta$. This
implies that $m_1$ is also symmetric under this reflection and thus
depends only on $\cos \delta$ (not $\sin\delta$). As a result, even
though there are two solutions for $\delta$, $m_1$ (and thus $m_{2,3}$
as well) is uniquely determined. Finally, by substituting the above
results into Eq.~\eqref{eq:alp23}, we determine the Majorana CP phases
$\alpha_2$ and $\alpha_3$. Again, since $R^*_{2,3}(-\delta) = R_{2,3}
(\delta)$, we have $\alpha_{2,3} (-\delta) = -\alpha_{2,3}(\delta)$, as
seen in Fig.~\ref{fig:alpha} in Appendix~\ref{app:r2r3}.

In the next section, we discuss the predictions of the minimal gauged
U(1)$_{L_\mu- L_\tau}$ model with the recent oscillation data. Before
closing this section, let us give some general remarks. 
\begin{itemize}
\item
If the U(1)$_{L_\mu- L_\tau}$ symmetry breaking scale is much higher
than the electroweak scale, we expect sizable quantum corrections to the
neutrino mass matrix. Such quantum corrections can be taken into account
by using renormalization group equations. Remarkably, it is found that the
two-zero minor structure of ${\cal M}_{\nu_L}$ is preserved throughout
the renormalization group flow \cite{Lavoura:2004tu, Lashin:2007dm}. To
see this, we first 
note that below the right-handed neutrino mass scale, which is around
the U(1)$_{L_\mu- L_\tau}$ symmetry breaking scale in this model,
right-handed neutrinos are integrated out to give the following
dimension-five effective operator:
\begin{equation}
 {\cal L}_{\rm eff} = \frac{1}{2}C_{\alpha \beta} (L_\alpha \cdot H)
(L_\beta \cdot H) +\text{h.c.}~,
\end{equation}
where $C_{\alpha\beta}$ has the two-zero minor structure at the
right-handed neutrino mass scale. The renormalization group equation of
the Wilson coefficient $C_{\alpha\beta}$ at one-loop level is
\cite{Antusch:2001ck} 
\begin{equation}
 \mu \frac{dC}{d\mu} = -\frac{3}{32\pi^2}\left[
\left(Y_e^\dagger Y_e^{}\right)^T C +
C \left(Y_e^\dagger Y_e^{}\right)
\right]+\frac{K}{16\pi^2} C~,
\end{equation}
with 
\begin{equation}
 K = -3g_2^2 + 2 {\rm Tr} \left(3 Y_u^\dagger Y_u^{} + 3Y_d^\dagger
Y_d^{} + Y_e^\dagger Y_e^{}
\right) + 2\lambda ~,
\end{equation}
where $Y_u$, $Y_d$, and $Y_e$ denote the up-type, down-type, and
charged-lepton Yukawa matrices, respectively, $g_2$ is the SU(2)$_L$
gauge coupling, and $\lambda$ is the Higgs quartic coupling: ${\cal
L}_{\rm quart} = -\frac{1}{2}\lambda (H^\dagger H)^2$. Now recall that
the charged lepton Yukawa matrix is diagonal in our model. In this case,
the above equation can readily be solved as follows \cite{Ellis:1999my}: 
\begin{equation}
 C(t) = I_K (t) \, {\cal I}(t)\, C(0) \,{\cal I}(t)~,
\label{eq:ctreltoc0}
\end{equation}
where $t \equiv \ln(\mu/\mu_0)$ with $\mu_0$ being the initial scale,
and 
\begin{align}
 I_K(t) &= \exp \left[
\frac{1}{16\pi^2} \int_0^t K(t^\prime)\, dt^\prime 
\right]~,
\qquad
{\cal I}(t) =\exp \left[-
\frac{3}{32\pi^2} \int_0^t Y^\dagger_e Y_e^{}(t^\prime)\, dt^\prime 
\right]~.
\label{eq:intdef}
\end{align}
Note that ${\cal I}(t)$ is a diagonal matrix. Therefore, if $C_{\mu
\mu}^{-1} (0) = C^{-1}_{\tau \tau} (0) = 0$, then $C_{\mu
\mu}^{-1} (t) = C^{-1}_{\tau \tau} (t) = 0$, which proves that
the two-zero minor structure of the Wilson coefficient $C$ remains
at low energies. As a result, the two-zero minor neutrino-mass structure
in our model is robust against quantum corrections, even if the
U(1)$_{L_\mu- L_\tau}$ symmetry breaking scale is much higher than the
electroweak scale.

\item
By performing a similar analysis, we can study the minimal gauged
U(1)$_{L_e- L_\mu}$ and U(1)$_{L_e- L_\tau}$ models. We again obtain two
conditions similar to Eqs.~\eqref{eq:vanmu} and \eqref{eq:vantau}, which
follow from the zero components in the inverse of the neutrino mass
matrix in each model. We however find that in these cases it is unable
to find parameter space consistent with the observed values of the
neutrino oscillation parameters, as demonstrated in
Appendix~\ref{sec:othercases}. We thus conclude that within the minimal
approach discussed in this paper, the U(1)$_{L_\mu- L_\tau}$ case is the
only possibility that offers a desirable neutrino mass structure. 

\item
If we go beyond the minimal model, we may obtain a different
neutrino-mass structure in the presence of the U(1)$_{L_\mu- L_\tau}$
gauge symmetry. For example, Refs.~\cite{Baek:2015mna, Lee:2017ekw}
discuss gauged U(1)$_{L_\mu- L_\tau}$ models where neutrino masses are
radiatively induced at one-loop level. It turns out that the neutrino
mass matrices in these models have the form of the two-zero texture
\cite{Berger:2000zj, Frampton:2002yf, Xing:2002ta, Kageyama:2002zw,
Xing:2002ap}, which predicts IO for neutrino mass
spectrum. Thus, the determination of neutrino mass hierarchy in future
neutrino experiments allows us to distinguish this class of models from
the minimal model, where NO is predicted as we have seen above. 

\end{itemize}

\section{Predictions for the neutrino parameters}
\label{sec:result}

\begin{table}[t]
 \begin{center}
\caption{Input values for the neutrino oscillation parameters we use
  in this paper. We take them from Ref.~\cite{Capozzi:2017ipn}. }
\label{tab:input}
\vspace{5pt}
\begin{tabular}{lccc}
\hline
\hline
 Parameter\qquad & Best fit \qquad& 1$\sigma$ range \qquad& 2$\sigma$ range \\
\hline
$\delta m^2/10^{-5}~\text{eV}^2$ & 7.37 & 7.21--7.54 & 7.07--7.73 \\
$\Delta m^2/10^{-3}~\text{eV}^2$ & 2.525 & 2.495--2.567 &2.454--2.606 \\
$\sin^2 \theta_{12}/10^{-1}$ & 2.97 & 2.81--3.14 & 2.65--3.34 \\
$\sin^2 \theta_{23}/10^{-1}$ & 4.25 & 4.10--4.46 & 3.95--4.70 \\
$\sin^2 \theta_{13}/10^{-2}$ & 2.15 & 2.08--2.22 & 1.99--2.31 \\
$\delta/\pi$ & 1.38 & 1.18--1.61 & 1.00--1.90 \\
\hline
\hline
\end{tabular}
 \end{center}
\end{table}

Using the results obtained above, we now compute quantities relevant to
neutrino experiments with the errors in the neutrino
oscillation parameters taken into account. For input values, we use the
values given in Ref.~\cite{Capozzi:2017ipn}, which are summarized in
Table~\ref{tab:input}.  In particular, we take the three mixing angles
and the two mass squared differences ,
\begin{align}
\theta_{12},~ \theta_{23},~ \theta_{13},~ \delta m^2,~ \Delta m^2~,
\end{align}
as input parameters, and evaluate the predicted values of the other
parameters, including Dirac CP phase $\delta$, the absolute masses
$m_i$, their sum $\sum_i m_i$, and the effective Majorana neutrino mass
$\langle m_{\beta\beta} \rangle$. The prediction for the Majorana phases
$\alpha_2$ and $\alpha_3$ is also presented in Appendix~\ref{app:r2r3}.

\begin{figure}[t]
\centering
\includegraphics[clip, width = 0.65 \textwidth]{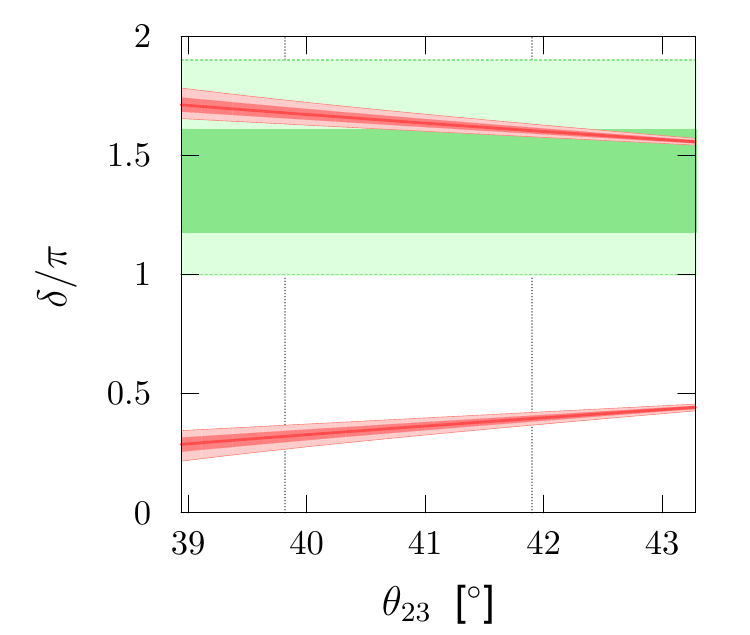}
\caption{The prediction for the Dirac CP phase $\delta$ in the minimal
 gauged U(1)$_{L_\mu- L_\tau}$ model. The red lines show the CP phase
 $\delta$ against $\theta_{23}$. $\theta_{23}$ is varied in the
 2$\sigma$ range, and the 1$\sigma$ range is in between the vertical
 thin dotted lines. The dark (light) red bands show the uncertainty
 coming from the 1$\sigma$ ($2\sigma$) errors in the parameters
 $\theta_{12}$, $\theta_{13}$, $\delta m^2$, and $\Delta m^2$. We also
 show the 1$\sigma$ (2$\sigma$) favored region of $\delta$ in the dark
 (light) horizontal green bands.} 
\label{fig:delta}
\end{figure}

In Fig.~\ref{fig:delta}, we plot the Dirac CP phase $\delta$ as
functions of $\theta_{23}$ in the red lines. We vary $\theta_{23}$ in
the 2$\sigma$ renege, where the 1$\sigma$ range is in between the
vertical thin dotted lines. The dark (light) red bands show the
uncertainty coming from the 1$\sigma$ ($2\sigma$) errors in the other
parameters $\theta_{12}$, $\theta_{13}$, $\delta m^2$, and $\Delta m^2$.
We find that this uncertainty is
dominated by the error in $\theta_{12}$. We also show the 1$\sigma$
(2$\sigma$) favored region of $\delta$ in the dark (light) horizontal green
bands. As we discussed in the previous section, there are two solutions
for $\delta$ for each value of $\theta_{23}$. Intriguingly, the upper
line is right in the middle of the favored range of $\delta$; in
particular, $\theta_{23} \simeq 41.5^\circ$ gives $\delta \simeq
1.6\pi$, both of which are within the $1\sigma$ allowed
region. Consequently, this model predicts $\delta \simeq
1.59\pi$--$1.70\pi$ ($1.54\pi$--$1.78\pi$) within $1\sigma$
($2\sigma$). Future neutrino experiments can test this prediction
through precision measurements of $\theta_{23}$ and $\delta$
\cite{deGouvea:2013onf}.

\begin{figure}[t]
  \centering
  \subcaptionbox{\label{fig:mass} Mass spectrum}{
  \includegraphics[width=0.48\columnwidth]{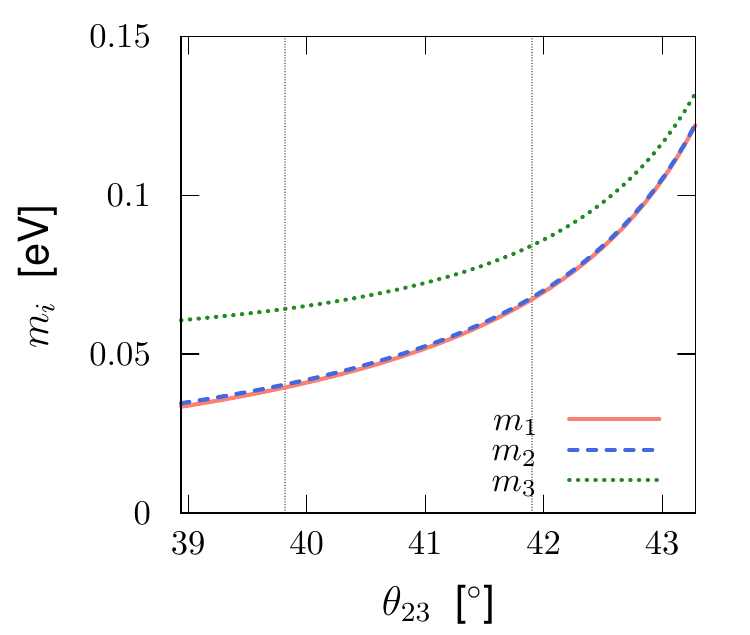}}
  \subcaptionbox{\label{fig:sum} $\sum_{i} m_i$}{
  \includegraphics[width=0.48\columnwidth]{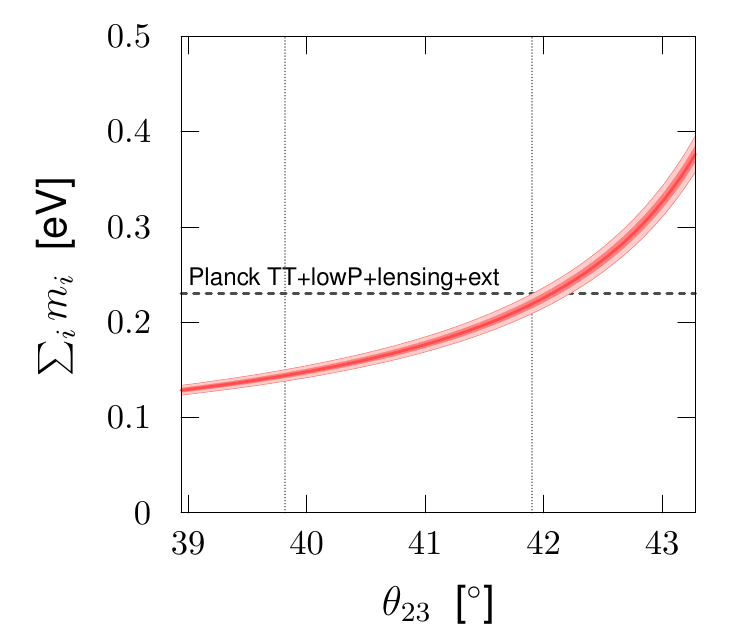}}
\caption{(a) The prediction for the neutrino masses in the minimal
 gauged U(1)$_{L_\mu- L_\tau}$ model. The neutrino masses $m_i$ are
 shown as functions of $\theta_{23}$. The other four parameters
 ($\theta_{12}$, $\theta_{13}$, $\delta m^2$, and $\Delta m^2$) are
 fixed to their best-fit values. (b) The sum of the neutrino masses as a
 function of $\theta_{23}$. The dark (light) red band shows the
 uncertainty coming from the 1$\sigma$ ($2\sigma$) errors in the
 parameters $\theta_{12}$, $\theta_{13}$, $\delta m^2$, and $\Delta
 m^2$. The entire region is within the 2$\sigma$ range of $\theta_{23}$,
 while its 1$\sigma$ range is between the thin vertical dotted lines. We
 also show in the black dashed line the present limit imposed by the
 Planck experiment: $\sum_{i} m_i < 0.23$~eV (Planck TT+lowP+lensing+ext)
 \cite{Ade:2015xua}.} 
  \label{fig:mi}
\end{figure}

Next, we evaluate the neutrino masses $m_i$, which are shown in
Fig.~\ref{fig:mass} as functions of $\theta_{23}$. Here, the other
parameters are fixed to be their best-fit values. We see that all of
these masses are predicted to be $\gtrsim \sqrt{\Delta m^2} \simeq
5\times 10^{-2}$~eV. We also plot the sum of these neutrino masses as a
function of $\theta_{23}$ in Fig.~\ref{fig:sum}, where the dark
(light) red band shows the uncertainty coming from the 1$\sigma$
($2\sigma$) errors in the parameters other than $\theta_{23}$. In this
case, it turns out that the dominant contribution to the uncertainty
(except for that from the error in $\theta_{23}$) comes from the error
in $\theta_{13}$, though the error in $\Delta m^2$ also gives a sizable
contribution. We also show in the black dashed line the present limit
imposed by the Planck experiment: $\sum_{i} m_i < 0.23$~eV (Planck
TT+lowP+lensing+ext) \cite{Ade:2015xua}.\footnote{If we exclude the
Planck lensing data, we obtain a slightly stringent bound: $\sum_{i} m_i
< 0.17$~eV \cite{Ade:2015xua}.} From this figure, we find that a wide
range of the parameter region predicts a value of $\sum_{i}m_i$ which is
below the present limit, though a part of the parameter region has
already been disfavored by the Planck limit. We however note that this
bound relies on the standard cosmological history, and thus if some new
physics effects modify the cosmological evolution, then this bound may
significantly be relaxed. In any case, our model predicts a rather large
value of the sum of the neutrino masses, $\sum_{i} m_i \simeq
0.14$--0.22~eV (0.12--0.40~eV) at $1\sigma$ ($2\sigma$), which may be
probed in future cosmological observations.

\begin{figure}[t]
\centering
\includegraphics[clip, width = 0.65 \textwidth]{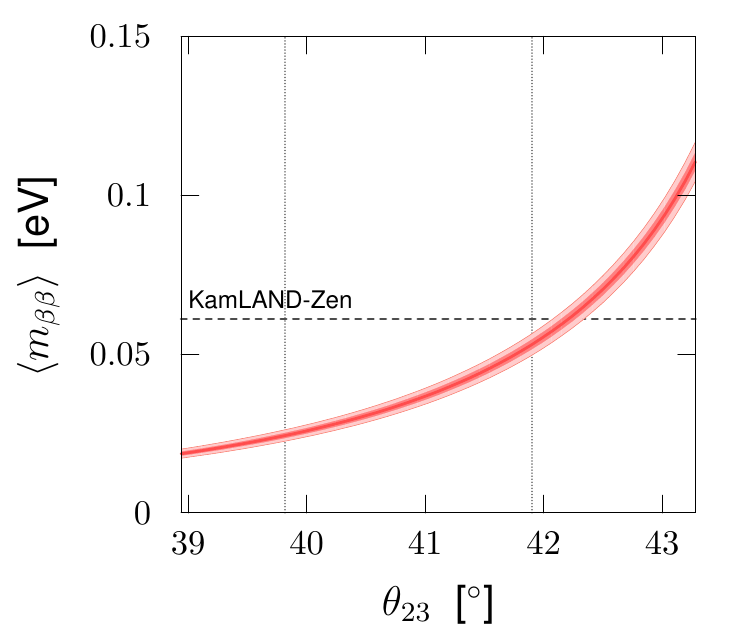}
\caption{The prediction for the effective Majorana neutrino mass
$\langle m_{\beta\beta} \rangle$ as a function of
 $\theta_{23}$ in the minimal gauged U(1)$_{L_\mu- L_\tau}$ model.
The dark (light) red band shows the uncertainty coming
 from the 1$\sigma$ ($2\sigma$) errors in the parameters other than
 $\theta_{23}$. The entire region is within the 2$\sigma$ range of
 $\theta_{23}$, while its 1$\sigma$ range is between the thin vertical
 dotted lines.  We also show the strongest limit from KamLAND-Zen,
 $\langle m_{\beta\beta} \rangle <0.061$~eV, in the black dashed line
 \cite{KamLAND-Zen:2016pfg}.  }
\label{fig:mbb}
\end{figure}

These relatively large values of $m_i$ open up a possibility of
testing this model in neutrinoless double-beta decay experiments. 
The rate of neutrinoless double-beta decay is proportional to the square
of the effective Majorana neutrino mass $\langle m_{\beta\beta}
\rangle$, which is given by
\begin{align}
 \langle m_{\beta\beta} \rangle \equiv \left|
\sum_{i} U_{ei}^2 m_i
\right|
= \left|
c_{12}^2 c_{13}^2 m_1 +
s_{12}^2 c_{13}^2 e^{i \alpha_2} m_2 +
s_{13}^2 e^{i(\alpha_3 -2\delta)} m_3 
\right| ~.
\end{align}
It should be emphasized that, in the minimal gauged U(1)$_{L_\mu-
L_\tau}$ model, not only the neutrino masses $m_i$ but also the Majorana
phases $\alpha_{2,3}$ are uniquely determined as functions of the other
neutrino oscillation parameters. Thus, the value of the effective
mass $\langle m_{\beta\beta} \rangle$ is also predicted
unambiguously. Note also that this quantity has reflection symmetry with
respect to $\delta \to -\delta$ and thus depends only on $\cos \delta$. 
In Fig.~\ref{fig:mbb}, we show 
$\langle m_{\beta\beta} \rangle$ as a function of $\theta_{23}$, where 
the dark (light) red band shows the uncertainty coming from the 1$\sigma$
($2\sigma$) errors in the parameters other than
$\theta_{23}$. Currently, the KamLAND-Zen experiment gives the strongest
bound on $\langle m_{\beta\beta} \rangle$: $\langle m_{\beta\beta}
\rangle <0.061$--0.165~eV \cite{KamLAND-Zen:2016pfg} where the
uncertainty stems from the estimation of the nuclear matrix element for
$^{136}$Xe. We also show in Fig.~\ref{fig:mbb} the most severe bound
from KamLAND-Zen, $\langle m_{\beta\beta} \rangle <0.061$~eV, in the
black dashed line \cite{KamLAND-Zen:2016pfg}. As can be seen, most
parameter region predicts a value of $\langle m_{\beta\beta} \rangle$
lower than the strongest bound. At $1\sigma$ ($2\sigma$) level, this model
predicts $\langle m_{\beta \beta }\rangle \gtrsim 0.024$~eV
(0.017~eV)---this can be within the reach of future neutrinoless
double-beta decay experiments \cite{Vergados:2016hso}.

\section{Implications for Leptogenesis}
\label{sec:LG}

In this section, we discuss the implications of our results for the leptogenesis scenario~\cite{Fukugita:1986hr}, which is one of the most attractive mechanisms to explain the origin of the baryon asymmetry of the Universe.
The minimal gauged U(1)$_{L_\mu-L_\tau}$ model has three right-handed neutrinos coupled to the Standard Model leptons, 
and therefore it contains enough ingredients for the leptogenesis: the CP-violating decay of the right-handed neutrino in the early universe can generate the lepton asymmetry, which is converted to the baryon asymmetry via the sphaleron process~\cite{Kuzmin:1985mm}.

As we have seen in the previous sections, the light neutrino mass matrix ${\cal M}_{\nu_L} \simeq U^* \text{diag}(m_1, m_2, m_3) U^\dagger$ in the minimal gauged U(1)$_{L_\mu-L_\tau}$ model is uniquely determined for a given set of neutrino oscillation parameters $\theta_{12},~ \theta_{23},~ \theta_{13},~ \delta m^2,~ \Delta m^2$, and $\text{sign}(\sin\delta)$. Therefore, the mass matrix of the right-handed neutrino, ${\cal M}_R \simeq - {\cal M}_D  {\cal M}_{\nu_L}^{-1} {\cal M}_D$, 
is also tightly constrained, having only three additional free parameters, $\lambda_{e}$, $\lambda_\mu$ and $\lambda_\tau$. Note that these neutrino Yukawa couplings can be taken to be real and positive by field redefinitions. Therefore, there is no additional phase parameter in the model.
By diagonalizing the masses of the right-handed neutrinos, we can rewrite the Lagrangian \eqref{eq:lag} as
\begin{align}
 \Delta {\cal L} = 
&
-\sum_{i=1}^3 \sum_{\alpha=e,\mu,\tau}
\widehat{\lambda}_{i\alpha} \widehat{N}_i^c (L_\alpha \cdot H)
-\frac{1}{2}\sum_{i=1}^3M_i \widehat{N}_i^c \widehat{N}_i^c +\text{h.c.} ~,
\label{eq:lag2}
\end{align}
where $\widehat{N}_i^c$ are the right-handed neutrino fields in the basis where the masses are diagonalized, $\widehat{\lambda}_{i\alpha}$ are the neutrino Yukawa couplings in that basis, and $M_i$ are the masses which are taken to be real and positive. Explicitly, they are given by
\begin{align}
{\cal M}_R &= \Omega^* \text{diag}(M_1, M_2, M_3) \Omega^\dagger, \quad \Omega^\dagger \Omega = I,
\\
\widehat{N}^c_i &= \sum_{\alpha} \Omega_{\alpha i}^* N_\alpha^c,
\\
\widehat{\lambda}_{i\alpha} &= \Omega_{\alpha i}\lambda_{\alpha}~~\text{(not summed)}.
\end{align}
We emphasize again that both masses $M_i$ and the couplings $\widehat{\lambda}_{i\alpha}$ are completely determined by the three real parameters $\lambda_{e,\mu,\tau}$ and the oscillation parameters $\theta_{12},~ \theta_{23},~ \theta_{13},~ \delta m^2,~ \Delta m^2$,  $\text{sign}(\sin\delta)$. 

One of the most important parameters in the leptogenesis scenario is the asymmetry parameter $\epsilon_1$, which represents the lepton asymmetry generated by the decay of the lightest right-handed neutrino. At the leading order, it is given by~\cite{Flanz:1994yx,Covi:1996wh,Buchmuller:1997yu}
\begin{align}
\epsilon_1 &= \frac{1}{8\pi}
\frac{1}{ (\widehat{\lambda}\widehat{\lambda}^\dagger)_{11} }
\sum_{j=2,3}
\text{Im}\left[
\{(\widehat{\lambda}\widehat{\lambda}^\dagger)_{1j}\}^2
\right]
f\left(\frac{M_j^2}{M_1^2}\right),
\label{eq:eps1}
\\
f(x) &= \sqrt{x}\left[
1-(x+1)\ln\left(1+\frac{1}{x}\right)
-\frac{1}{x-1}
\right].
\label{eq:fx}
\end{align}
Here, one can see that there is a correlation between the sign of the Dirac phase $\delta$ and the baryon asymmetry of the Universe. Suppose that the sign of the Dirac phase $\delta$ is flipped, $\delta\to -\delta$, while the other input oscillation parameters $\theta_{12},~ \theta_{23},~ \theta_{13},~ \delta m^2,~ \Delta m^2$ as well as the neutrino Yukawa couplings $\lambda_{e,\mu,\tau}$ are fixed. As discussed in the previous sections, the Majorana phases then flip the sign, $\alpha_{2,3}\to -\alpha_{2,3}$, while the absolute masses of light neutrinos do not change, $m_i\to m_i$. Thus, the PMNS matrix transforms as $U\to U^*$. This then results in ${\cal M}_{\nu_L}\to {\cal M}_{\nu_L}^*$, leading to ${\cal M}_R\to {\cal M}_R^*$, $\Omega\to \Omega^*$, $\widehat{\lambda}\to \widehat{\lambda}^*$, and eventually $\epsilon_1\to -\epsilon_1$.
This means that, for a given input oscillation parameters $\theta_{12},~ \theta_{23},~ \theta_{13},~ \delta m^2,~ \Delta m^2$ and the neutrino Yukawa couplings $\lambda_{e,\mu,\tau}$, there is one-to-one correspondence between the sign of the Dirac phase $\delta$ and the sign of the baryon asymmetry of the Universe.

\begin{figure}[t]
\centering
\includegraphics[clip, width = 0.6 \textwidth]{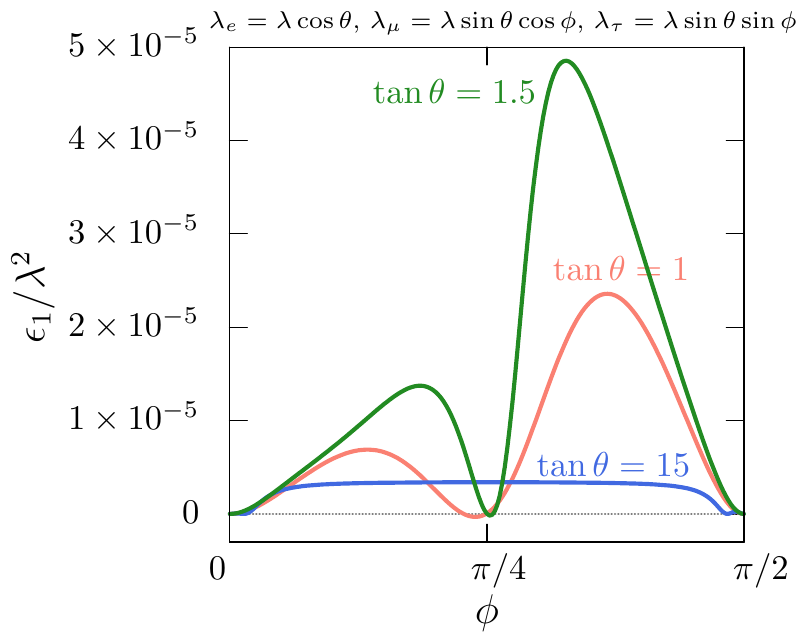}
\caption{The asymmetry parameter $\epsilon_1/\lambda^2$ as a function of the diagonal neutrino Yukawa couplings, which are parametrized as 
$(\lambda_e,\lambda_\mu,\lambda_\tau)
=\lambda(\cos\theta, \sin\theta\cos\phi, \sin\theta\sin\phi)$. The input oscillation parameters $(\theta_{12},  \theta_{23},  \theta_{13}, \delta m^2, \Delta m^2)$ are taken to be their best-fit values in Table~\ref{tab:input}, while the sign of the Dirac phase $\delta$ is taken to be negative, $\delta<0$ (or $\delta>\pi$).}
\label{fig:eps1}
\end{figure}

\begin{figure}[t]
\centering
\includegraphics[clip, width = 0.45 \textwidth]{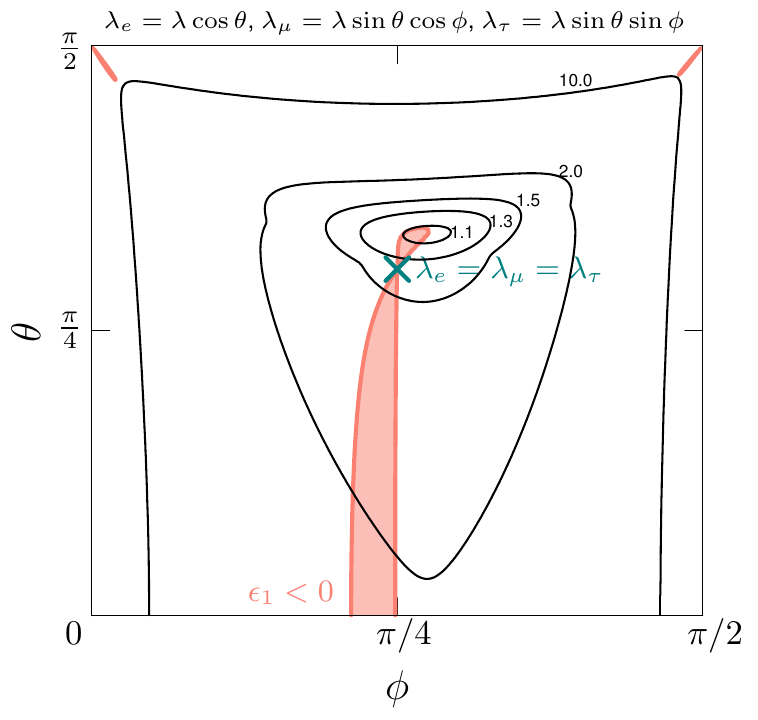}
\includegraphics[clip, width = 0.45 \textwidth]{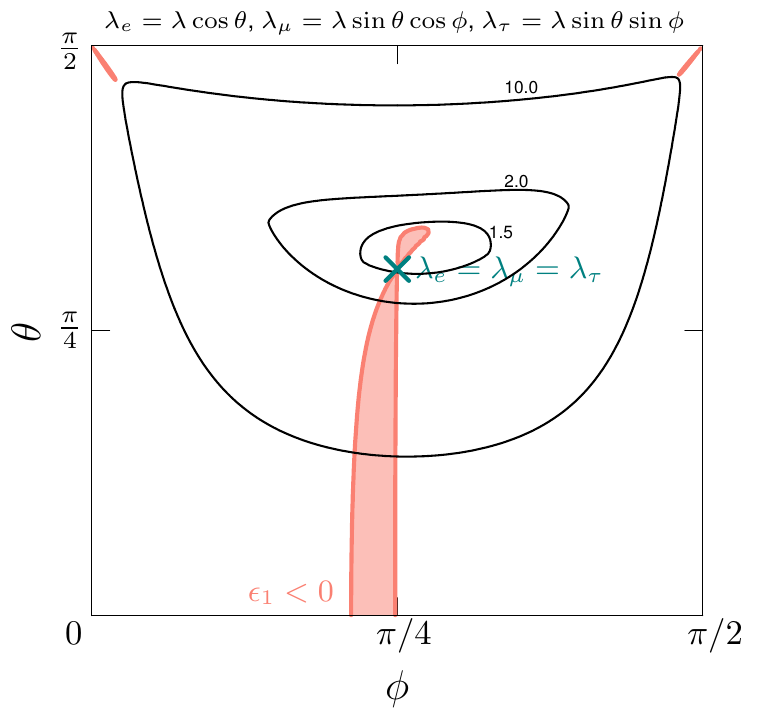}
\caption{The sign of the asymmetry parameter $\epsilon_1$ and contours of the right-handed neutrino mass ratios $M_2/M_1$ (left) and $M_3/M_1$ (right) in the $(\theta, \phi)$ plane, where the neutrino Yukawa couplings are parametrized as 
$(\lambda_e,\lambda_\mu,\lambda_\tau)
=\lambda(\cos\theta, \sin\theta\cos\phi, \sin\theta\sin\phi)$. The input
 oscillation parameters $(\theta_{12},  \theta_{23},  \theta_{13},
 \delta m^2, \Delta m^2)$ are taken to be their best-fit values in
 Table~\ref{tab:input}, while the sign of the Dirac phase $\delta$ is
 taken to be negative, $\delta<0$ (or $\delta>\pi$). The green cross
 corresponds to $\lambda_e = \lambda_\mu = \lambda_\tau$. }
\label{fig:signANDm1m23}
\end{figure}

In Fig.~\ref{fig:eps1}, we show the asymmetry parameter
$\epsilon_1/\lambda^2$ as a function of the neutrino Yukawa couplings, which are parametrized as 
$(\lambda_e,\lambda_\mu,\lambda_\tau)
=\lambda(\cos\theta, \sin\theta\cos\phi,
\sin\theta\sin\phi)$.\footnote{We can take account of the one-loop
renormalization-group effects by instead regarding $I_K^{1/2}(t)
\lambda_\alpha {\cal I}_{\alpha \alpha}$ ($\alpha = e, \mu, \tau$) as
input parameters (see Eqs.~\eqref{eq:ctreltoc0} and
\eqref{eq:intdef}). Thus, our consequence here is robust against
one-loop quantum corrections. } 
Note that the parameter $\epsilon_1$ scales as $\epsilon_1\propto \lambda^2$ in this parametrization, as shown in Eq.~\eqref{eq:eps1}. The oscillation parameters $\theta_{12},~ \theta_{23},~ \theta_{13},~ \delta m^2,~ \Delta m^2$
are taken to be their best-fit values in Table~\ref{tab:input}. The sign of the Dirac phase $\delta$ is taken to be negative, $\delta<0$ (or $\delta>\pi$), as it is favored at 2$\sigma$ level.
Note that the observed baryon asymmetry of the Universe requires $\epsilon_1<0$, because the sphaleron process predicts $n_B/n_L <0$~\cite{Harvey:1990qw}. Surprisingly, negative $\epsilon_1$ is realized only in the limited regions of the parameter space.
This is more clearly seen in Figs.~\ref{fig:signANDm1m23}, where we show
the regions of $\epsilon_1<0$ in the $(\theta,\phi)$ plane as red shaded
areas. We also show
the contours of the right-handed neutrino mass ratios $M_{2}/M_1$ and
$M_3/M_1$ in the left and right panels, respectively. As we see, the
asymmetry parameter $\epsilon_1$ can be negative when some of the
right-handed neutrinos are degenerate in mass---$M_1 \simeq M_2$ in the
negative $\epsilon_1$ region around $\phi \simeq \pi/4$ while $M_2
\simeq M_3$ for $\theta \simeq \pi/2$ and $\phi \simeq 0, \pi/2$. 
In the latter regions, $\theta \simeq \pi/2$ and $\phi \simeq 0, \pi/2$,
the lightest right-handed neutrino mass is much smaller than the other
ones, and $\lambda_{e, \tau} \ll \lambda_\mu$ ($\lambda_{e, \mu} \ll
\lambda_\tau$) for $\phi \simeq 0$ ($\phi \simeq \pi/2$). In these
cases, the absolute value of the asymmetry parameter $|\epsilon_1|$ is
found to be quite suppressed, and thus it is rather difficult to obtain
a sizable value, say $|\epsilon_1| \gtrsim 10^{-6}$.\footnote{Note that
the function $f(x)$ in Eq.~\eqref{eq:fx} goes as $f(x) \simeq
-3/(2\sqrt{x})$ for $x\gg 1$. } Similarly, $|\epsilon_1|$ gets small for
$\phi \simeq \pi/4$ and $\theta \ll \pi/4$. These observations further
restrict the promising parameter region for leptogenesis to be
$(\theta, \phi) \simeq (\pi/4, \pi/4)$.

The final baryon asymmetry depends on the production mechanism of the right-handed neutrino.
In the case of thermal leptogenesis, the predicted range of the lightest neutrino mass in the present model, $m_1\gtrsim 0.03~\text{eV}$, corresponds to the so-called strong wash-out region (see, e.g.,~\cite{Buchmuller:2004nz}).
Moreover, as we see from Fig.~\ref{fig:signANDm1m23}, around $(\theta, \phi)
\simeq (\pi/4, \pi/4)$, right-handed neutrino masses are close to each
other. In this case, to properly estimate the net baryon asymmetry
generated via the decay of right-handed neutrinos, we may need to include
not only the contribution of the lightest right-handed neutrino but
those of the other ones. 
Flavor effects~\cite{Abada:2006fw,Nardi:2006fx,Abada:2006ea} may also become important.
A more detailed study will be given elsewhere~\cite{AHN_prep}.

\section{Conclusions}
\label{sec:conclusion}

We have studied the structure of the neutrino mass matrix in the minimal
gauged U(1)$_{L_\mu-L_\tau}$ model. Because of the U(1)$_{L_\mu-L_\tau}$
gauge symmetry, the structure of both Dirac and Majorana mass terms of
neutrinos is tightly restricted, which results in a two-zero minor
structure of the neutrino mass matrix. Because of this restriction, all
the CP phases and the neutrino masses are uniquely determined. We find
that this model gives the following prediction at $1\sigma$ ($2\sigma$)
level: 
\begin{itemize}
\item Dirac CP phase:  $\delta\simeq 1.59\pi$--$1.70\pi$
      ($1.54\pi$--$1.78\pi$), 
\item the sum of the neutrino masses: $\sum_{i}m_i \simeq 0.14$--0.22~eV
      ($0.12$--$0.40$~eV),
\item the effective mass for the neutrinoless double beta decay:
$\langle m_{\beta \beta }\rangle \simeq 0.024$--0.055~eV
      (0.017--0.12~eV). 
\end{itemize}
They are totally consistent with
the current experimental limits, and hold independently of the
U(1)$_{L_\mu-L_\tau}$ breaking scale and the Majorana mass scale. In
this sense, the above predictions are the generic features of the
minimal gauged U(1)$_{L_\mu-L_\tau}$ model.  
Remarkably, these predictions can be
tested in various neutrino experiments in the near future---regardless of
the scale of the U(1)$_{L_\mu - L_\tau}$ symmetry breaking---which we
believe sheds light on the gauge structure of physics beyond the
Standard Model. 

We have also discussed the implications of the minimal gauged
U(1)$_{L_\mu - L_\tau}$ model for the leptogenesis scenario, and found
that the correct sign of the baryon asymmetry of the Universe can be
obtained only in the limited regions of the parameter space, as the
right-handed neutrino mass structure is also severely restricted in this
model. In particular, the observed value of baryon asymmetry can be
realized when right-handed neutrinos are degenerate in mass, which
requires a further detailed study to assess the viability of
leptogenesis in this model~\cite{AHN_prep}.

\section*{Acknowledgments}
This work is supported by the Grant-in-Aid for Scientific Research
(No.26104001 [KH], No.26104009 [KH], No.26247038 [KH], No.26800123 [KH],
No.16H02189 [KH], No.17K14270 [NN]), and by World Premier International
Research Center Initiative (WPI Initiative), MEXT, Japan. 
The work of KA is supported by the Program for Leading Graduate Schools,
MEXT, Japan.

\newpage
\section*{Appendix}
\appendix

\section{Miscellaneous formulae}
\label{sec:misc}
\renewcommand{\theequation}{A.\arabic{equation}}
\setcounter{equation}{0}

Here we give formulae that are useful for the study of the neutrino mass
structure in the minimal gauged U(1)$_{L_\mu-L_\tau}$ model.

\subsection{Neutrino mass matrix ${\cal M}_{\nu_L}$}
\label{app:mnul}

The light neutrino mass matrix ${\cal M}_{\nu_L}$ in Eq.~\eqref{eq:mnul}
can be expressed in terms of the Lagrangian parameters in
Eq.~\eqref{eq:lag} as\footnote{We obtain a different result from that
given in Ref.~\cite{Biswas:2016yan}.}
\begin{align}
 {\cal M}_{\nu_L} &=
\frac{v^2}{2(M_{ee} M_{\mu\tau} - 2 \lambda_{e\mu}
 \lambda_{e\tau} \langle \sigma\rangle^2)}
\nonumber \\[5pt]
&\times
\begin{pmatrix}
 -\lambda_e^2 M_{\mu\tau}
&
 \lambda_e \lambda_\mu \lambda_{e\tau} \langle \sigma \rangle
&
 \lambda_e \lambda_\tau \lambda_{e\mu} \langle \sigma\rangle
\\[5pt]
\lambda_e \lambda_\mu \lambda_{e\tau} \langle \sigma \rangle
&
- \frac{\lambda_\mu^2 \lambda_{e\tau}^2 \langle \sigma \rangle^2}{
 M_{\mu\tau}} 
&
 \frac{\lambda_\mu \lambda_\tau (-M_{ee} M_{\mu\tau} +\lambda_{e\mu}
 \lambda_{e\tau} \langle \sigma\rangle^2)}
{M_{\mu\tau}} 
\\[5pt]
\lambda_e \lambda_\tau \lambda_{e\mu} \langle \sigma\rangle 
&
 \frac{\lambda_\mu \lambda_\tau (-M_{ee} M_{\mu\tau} +\lambda_{e\mu}
 \lambda_{e\tau} \langle \sigma\rangle^2)}
{M_{\mu\tau}} 
&
 -\frac{\lambda_\tau^2 \lambda_{e\mu}^2 \langle \sigma \rangle^2}
{M_{\mu\tau}} 
\end{pmatrix}
~.
\end{align}
The determinant of this mass matrix is given by
\begin{equation}
 \text{det}\left({\cal M}_{\nu_L}\right)
=\frac{\lambda_e^2 \lambda_\mu^2 \lambda_\tau^2 v^6}
{8 M_{\mu\tau} \left(M_{ee} M_{\mu\tau}
-2\lambda_{e\mu} \lambda_{e\tau} \langle \sigma\rangle^2 
\right)}~.
\end{equation}
We find that this determinant vanishes if and only if $\lambda_\nu = 0$
($\nu = e$, $\mu$, or $\tau$). In this case, the mass matrix ${\cal
M}_{\nu_L}$ is block-diagonal, and thus cannot reproduce the required
neutrino mixing angles.

\subsection{$R_2$ and $R_3$}
\label{app:r2r3}

The functions $R_2(\delta)$ and $R_3(\delta)$ defined in
Eqs.~\eqref{eq:r2} and \eqref{eq:r3}, respectively, are expressed in
terms of neutrino oscillation parameters as 
\begin{align}
 R_2(\delta) &= -\frac{2\sin^2\theta_{12}\cos 2\theta_{23}
+ \sin 2\theta_{12} \sin 2\theta_{23} \sin \theta_{13} e^{i\delta}}
{2\cos^2\theta_{12}\cos 2\theta_{23}
- \sin 2\theta_{12} \sin 2\theta_{23} \sin \theta_{13} e^{i\delta}} ~,
\label{eq:r2ex}
\\[3pt]
R_3(\delta) &= -
\frac{\sin\theta_{13} e^{2i\delta}
\left[
2\cos 2\theta_{12} \cos 2\theta_{23} \sin \theta_{13} 
-\sin 2\theta_{12} \sin 2\theta_{23} (e^{-i\delta}+\sin^2 \theta_{13} e^{i\delta})
\right]
}{\cos^2 \theta_{13}
\left[
2\cos^2\theta_{12}\cos 2\theta_{23} -\sin 2\theta_{12} \sin
 2\theta_{23} \sin \theta_{13} e^{i\delta}\right]}~.
\label{eq:r3ex}
\end{align}
Using Eq.~\eqref{eq:delm2} together with Eq.~\eqref{eq:r2ex}, we find  
\begin{equation}
 m_1 = \delta m\biggl[
\frac{4s^4_{12} \cos^2 2\theta_{23}
+ 4s^3_{12} c_{12}s_{13}  \sin 4\theta_{23} 
\cos \delta + s^2_{13}
 \sin^2 2\theta_{12} \sin^2 2\theta_{23} 
}
{2\left(
2\cos 2\theta_{12} \cos^2 2\theta_{23} -
s_{13}
\sin 2\theta_{12} \sin
4\theta_{23}  \cos \delta
\right)}
\biggr]^{\frac{1}{2}}
~.
\end{equation} 
This expression shows that $m_1$ depends only on $\cos \delta$ (not
$\sin\delta$), as we have argued in Sec.~\ref{sec:model}.

\begin{figure}[t]
\centering
\includegraphics[clip, width = 0.6 \textwidth]{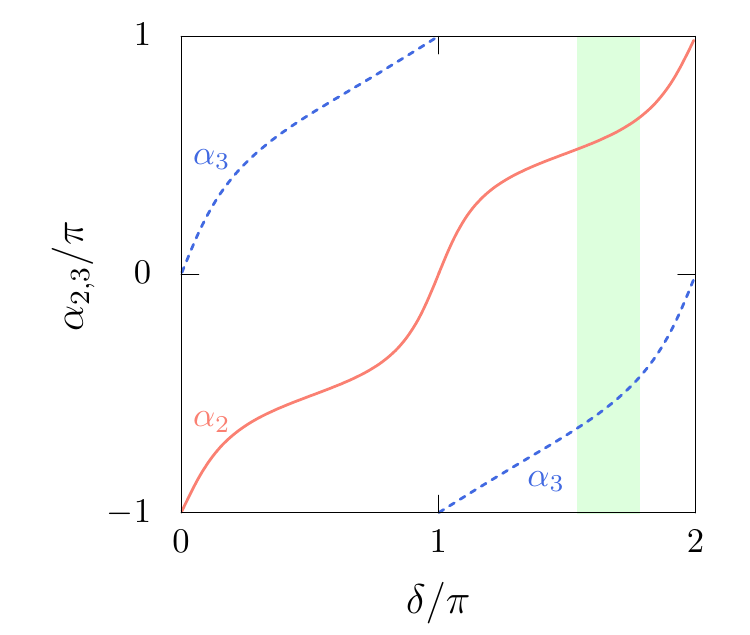}
\caption{Majorana phases $\alpha_i$ as functions of $\delta$. We fix the
 other neutrino oscillation parameters to be their best-fit values. The
 green band depicts the 2$\sigma$ favored region of $\delta$ predicted
 in the minimal gauged U(1)$_{L_\mu-L_\tau}$ model.}
\label{fig:alpha}
\end{figure}

Using Eqs.~\eqref{eq:alp23}, \eqref{eq:r2ex}, and \eqref{eq:r3ex}, we
determine the Majorana CP phases $\alpha_{2,3}$ as functions of
$\delta$, and show them in Fig.~\ref{fig:alpha}. Here, we fix the other
neutrino oscillation parameters to be their best-fit values. The green
band depicts the 2$\sigma$ favored region of $\delta$ predicted in the
minimal gauged U(1)$_{L_\mu-L_\tau}$ model. As can be seen, sizable
Majorana CP phases are predicted in this model. In addition, we find
$\alpha_{2,3} (-\delta) = -\alpha_{2,3}(\delta)$ (or $\alpha_{2,3} (\pi
-\delta) = -\alpha_{2,3}(\pi + \delta)$), as discussed in
Sec.~\ref{sec:model}.

\subsection{Cubic equation for $\cos \delta$}
\label{sec:cub}

Here, we show a cubic equation whose real solution in terms of $x$ gives
$\cos\delta$:  
\begin{align}
 & s^2_{13} \bigl[
4s^2_{13}\cos^2 2\theta_{12} \cos^2 2\theta_{23} 
-s_{13} \sin 4\theta_{12} \sin 4\theta_{23} (1+s_{13}^2) \,x
\nonumber \\
&+\sin^2 2\theta_{12} \sin^2 2\theta_{23} 
\left(c^4_{13} + 4s^2_{13} \,x^2\right)
\bigr]
\bigl[
2\left(
2\cos 2\theta_{12} \cos^2 2\theta_{23} -
s_{13}
\sin 2\theta_{12} \sin
4\theta_{23}  \,x
\right)
\bigr] \nonumber \\
&-\epsilon
\left[
4s^4_{12} \cos^2 2\theta_{23}
 + s^2_{13}
\sin^2 2\theta_{12} \sin^2 2\theta_{23}
+ 4s^3_{12} c_{12}s_{13}  \sin 4\theta_{23} \,
x
\right]  \nonumber \\
&\times\bigl[
4\cos^2 2\theta_{23} (c^4_{12} c^4_{13}-s^4_{13}\cos^2 2\theta_{12})
- s_{13}\sin 4\theta_{23}  
\{4c_{13}^4 c^3_{12} s_{12} -s^2_{13} \sin 4\theta_{12} (1+s^2_{13}) \}
\, x \nonumber \\ 
&- 4 s^4_{13} \sin^2 2 \theta_{12} \sin^2 2\theta_{23} \, x^2
\bigr] = 0 ~,
\label{eq:cubiceq}
\end{align}
where 
\begin{equation}
 \epsilon \equiv \frac{\delta m^2}{\Delta m^2 +{\delta m^2}/{2}}~.
\end{equation}
In the limit of $\epsilon \to 0$, the above equation leads to
\begin{equation}
\small
 4s^2_{13}\cos^2 2\theta_{12} \cos^2 2\theta_{23} 
-s_{13} \sin 4\theta_{12} \sin 4\theta_{23} (1+s_{13}^2) \,x
+\sin^2 2\theta_{12} \sin^2 2\theta_{23} 
\left(c^4_{13} + 4s^2_{13} \,x^2\right)=0 ~,
\label{eq:simpeq2}
\end{equation}
or
\begin{equation}
 2\cos 2\theta_{12} \cos^2 2\theta_{23} -
s_{13}
\sin 2\theta_{12} \sin
4\theta_{23}  \,x = 0~.
\label{eq:simpeq1}
\end{equation}
The discriminant of the quadratic equation \eqref{eq:simpeq2} is given
by 
\begin{equation}
 8c_{13}^4 s_{13}^2 \sin^2 2\theta_{12} \sin^2 2\theta_{23} 
(\cos 4\theta_{12} + \cos 4\theta_{23}) ~,
\end{equation}
which is negative as $\cos 4\theta_{12} + \cos 4\theta_{23} \simeq -1.63
< 0$. Thus, Eq.~\eqref{eq:simpeq2} does not give a real solution. On the
other hand, Eq.~\eqref{eq:simpeq1} gives 
\begin{equation}
 x =  \frac{\cot 2\theta_{12} \cot 2\theta_{23}}{\sin
  \theta_{13}} ~,
\label{eq:xsol}
\end{equation}
which agrees to Eq.~\eqref{eq:appcosd}. From the above derivation, we
see that the solution \eqref{eq:xsol} approximates the real solution of the
cubic equation \eqref{eq:cubiceq} with an accuracy of ${\cal
O}(\epsilon) = {\cal O}(\delta m^2/\Delta m^2)$.

\section{U(1)$_{L_e-L_\mu}$ and U(1)$_{L_e-L_\tau}$} 
\label{sec:othercases}
\renewcommand{\theequation}{B.\arabic{equation}}
\setcounter{equation}{0}

In this section, we examine the neutrino mass structure in the minimal
gauged U(1)$_{L_e-L_\mu}$ and U(1)$_{L_e-L_\tau}$ models and show that
it is unable to obtain a solution that is consistent with the observed
values of the neutrino oscillation parameters.\footnote{A similar
conclusion was also reached in Ref.~\cite{Liao:2013saa}. }

\subsection{U(1)$_{L_e-L_\mu}$}

\begin{figure}[t]
  \centering
  \subcaptionbox{\label{fig:emu} U(1)$_{L_e-L_\mu}$}{
  \includegraphics[width=0.48\columnwidth]{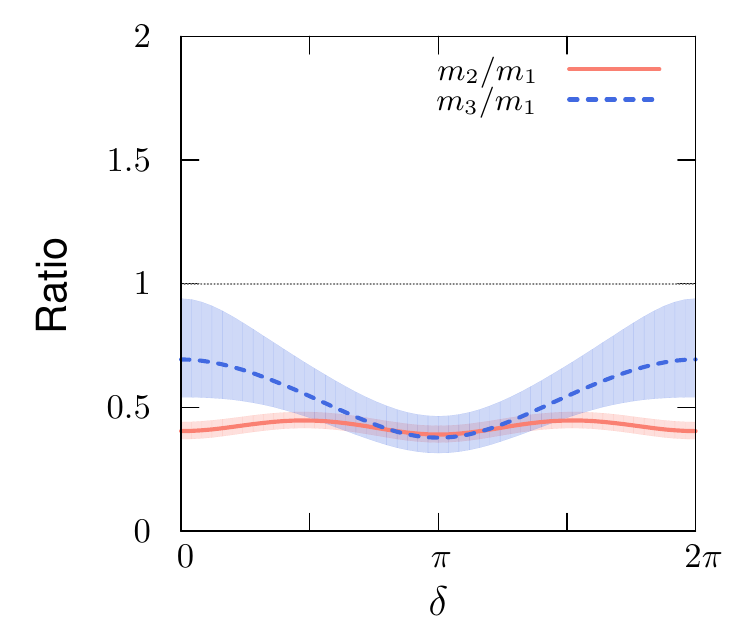}}
  \subcaptionbox{\label{fig:etau} U(1)$_{L_e-L_\tau}$}{
  \includegraphics[width=0.48\columnwidth]{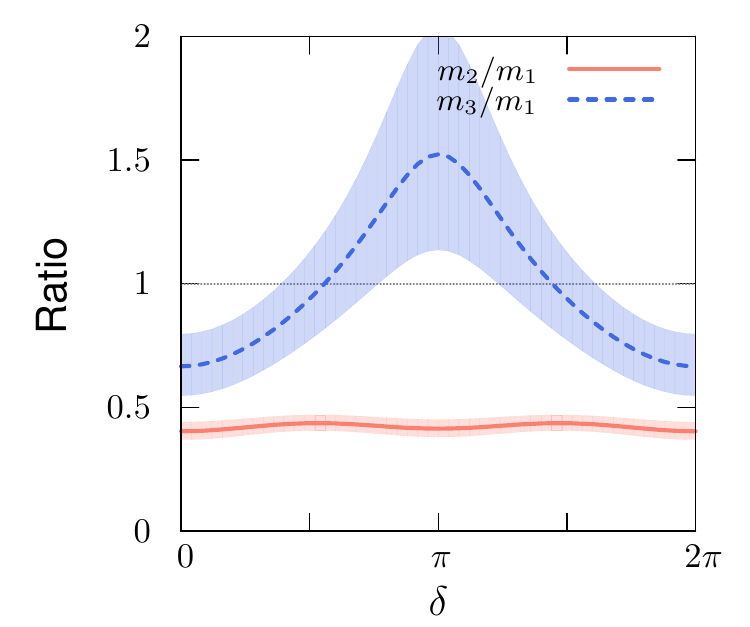}}
\caption{The mass ratios $m_2/m_1$ and $m_3/m_1$ as functions of the
 Dirac CP phase $\delta$ for the gauged (a) U(1)$_{L_e-L_\mu}$ and (b)
 U(1)$_{L_e-L_\tau}$ models. The bands show uncertainty coming from the
 1$\sigma$ error in the neutrino mixing parameters. The thin dotted line
 corresponds to $m_{2,3}/m_1 = 1$.} 
  \label{fig:emuetau}
\end{figure}

Following an analysis similar to that in Sec.~\ref{sec:model}, we find
that the $(e,e)$ and $(\mu, \mu)$ components in the inverse of the
neutrino mass matrix vanish in the minimal gauged U(1)$_{L_e-L_\mu}$
model. The corresponding two vanishing conditions are
\begin{align}
 \frac{1}{m_1}V_{e 1}^2 + \frac{1}{m_2}V_{e 2}^2\,e^{i\alpha_2}
+ \frac{1}{m_3}V_{e 3}^2\,e^{i\alpha_3} &= 0 ~,
\\[3pt]
  \frac{1}{m_1}V_{\mu 1}^2 + \frac{1}{m_2}V_{\mu 2}^2\,e^{i\alpha_2}
+ \frac{1}{m_3}V_{\mu 3}^2\,e^{i\alpha_3} &= 0 ~.
\end{align}
Solving these equations, we have
\begin{align}
 e^{i\alpha_2} &=\frac{m_2}{m_1} R_2^{e\mu} (\delta) ~,
\qquad
 e^{i\alpha_3} =\frac{m_3}{m_1} R_3^{e\mu} (\delta) ~,
\end{align}
with
\begin{align}
 R_2^{e\mu} &\equiv \frac{(V_{e 1} V_{\mu 3} + V_{e 3} V_{\mu
 1})V^*_{\tau 2}}
{(V_{e 2} V_{\mu 3}+ V_{e 3} V_{\mu 2}) V^*_{\tau 1}} ~, 
\\
 R_3^{e\mu} &\equiv \frac{(V_{e 1} V_{\mu 2} + V_{e 2} V_{\mu
 1})V^*_{\tau 3}}
{(V_{e 2} V_{\mu 3}+ V_{e 3} V_{\mu 2}) V^*_{\tau 1}} ~.
\end{align}
In Fig.~\ref{fig:emu}, we plot the mass ratios $m_2/m_1$ and $m_3/m_1$
as functions of $\delta$. As we see from this figure, $m_2 < m_1$ is
predicted for any value of $\delta$, and thus there is no solution which
gives an allowed pattern of neutrino mass spectrum.

\subsection{U(1)$_{L_e-L_\tau}$}

In this case, the $(e,e)$ and $(\tau, \tau)$ components in ${\cal
M}^{-1}_{\nu_L}$ are zero, which leads to
\begin{align}
 \frac{1}{m_1}V_{e 1}^2 + \frac{1}{m_2}V_{e 2}^2\,e^{i\alpha_2}
+ \frac{1}{m_3}V_{e 3}^2\,e^{i\alpha_3} &= 0 ~,
\\[3pt]
  \frac{1}{m_1}V_{\tau 1}^2 + \frac{1}{m_2}V_{\tau 2}^2\,e^{i\alpha_2}
+ \frac{1}{m_3}V_{\tau 3}^2\,e^{i\alpha_3} &= 0 ~.
\end{align}
These equations read
\begin{align}
 e^{i\alpha_2} &=\frac{m_2}{m_1} R_2^{e\tau} (\delta) ~,
\qquad
 e^{i\alpha_3} =\frac{m_3}{m_1} R_3^{e\tau} (\delta) ~,
\end{align}
with
\begin{align}
 R_2^{e\tau} &\equiv \frac{(V_{e 1} V_{\tau 3} + V_{e 3} V_{\tau
 1})V^*_{\mu 2}}
{(V_{e 2} V_{\tau 3}+ V_{e 3} V_{\tau 2}) V^*_{\mu 1}} ~, 
\\
 R_3^{e\tau} &\equiv \frac{(V_{e 1} V_{\tau 2} + V_{e 2} V_{\tau
 1})V^*_{\mu 3}}
{(V_{e 2} V_{\tau 3}+ V_{e 3} V_{\tau 2}) V^*_{\mu 1}} ~.
\end{align}
Using these equations, we plot the mass ratios $m_2/m_1$ and $m_3/m_1$
as functions of $\delta$ in Fig.~\ref{fig:etau}. Again, $m_2 < m_1$ over
the whole range of $\delta$, and thus this model cannot provide a
desirable neutrino mass ordering.

\newpage
{\small 
\bibliographystyle{JHEP}
\bibliography{ref}
}

\end{document}